\magnification=\magstep0
\hoffset=0.0 true cm
\voffset=0.0 true cm
\vsize=23.4 true cm
\hsize=17.0 true cm

\baselineskip=13pt
\parskip=10pt
\parindent=22pt

\raggedbottom
\nopagenumbers

\font\lilrm=cmr10 scaled \magstep1

\def\gsim{\vcenter{\hbox{$>$}\offinterlineskip\hbox{$\sim$}}}
\def\undertext#1{$\underline{\hbox{#1}}$}

\null
\centerline{\bf CMB OBSERVATIONS USING THE SKA}
\medskip
\centerline{\lilrm\bf {\undertext{Ravi Subrahmanyan}}$^{(1)}$, 
R.D. Ekers$^{(2)}$}
\medskip
\centerline{{\it $^{(1)}$Australia Telescope National Facility, CSIRO,}} 
\centerline{{\it Locked bag 194, Narrabri, NSW 2390, Australia.}}
\centerline{{\it E-mail:Ravi.Subrahmanyan@csiro.au}}
\medskip
\centerline{{\it $^{(2)}$Australia Telescope National Facility, CSIRO,}}
\centerline{{\it P O Box 76, Epping NSW 1710, Australia.}}
\centerline{{\it E-mail:Ron.Ekers@csiro.au}}
\bigskip
\rm

\noindent{\bf ABSTRACT}

\noindent We examine the prospects for observations of CMB anisotropy
with the SKA; we discuss the advantages of interferometric SKA imaging,
observing strategies, calibration issues and the achievable sensitivity. 
Although the SKA will probably operate at cm wavelengths, where discrete
source confusion dominates the CMB anisotropy,
its extreme sensitivity to point sources will make it possible to
subtract the source contamination at these
wavelengths and thereby image the low surface brightness CMB anisotropies 
on small angular scales.  
The SKA, operating at 10-20 GHz,  may usefully make 
high-$l$ observations of the CMB anisotropy spectrum and survey the sky 
for Sunyaev-Zeldovich decrements.

\noindent {\bf INTRODUCTION}

\noindent Observations of the angular anisotropy power in the Cosmic 
Microwave Background (CMB) temperature fluctuations
on different angular scales, leading to a measurement of the power spectrum
of the anisotropy, has proved to be very useful in determining the 
parameters of the space-time structure of the Universe, its constituents,
and understanding the ingredients and mechanism for large-scale 
structure growth and formation.  Current goals for CMB research include
extending the measurements of the anisotropy spectrum to small multipole
orders, measuring the CMB polarization anisotropy, and surveying the
sky for CMB decrements that have been created by cosmological clusters 
via the Sunyaev-Zeldovich effect (SZE).  The aims of these observations
are to provide additional constraints on cosmological parameters, break
parameter degeneracies, probe the cluster evolution and
deriving the equation of state for the dark energy.

\noindent The CMB anisotropy observations at high multipole orders, 
of the primary
anisotropy as well as of the SZE decrements towards cosmological clusters,
are currently being pursued with radio interferometers and bolometric arrays at
mm wavelengths.  Today, bolometers have the advantage of wider
bandwidths and better sensitivity particularly at the mm wavelengths
where most CMB telescopes choose to operate in order to avoid the
contamination and confusion arising from the non-thermal discrete
foreground sources in the sky.  However, interferometers have much lower
systematic errors and many ground based small angle anisotropy measurements 
are being done with purpose built interferometers operating at wavelengths 
around 10 mm. Because recent developments in low noise amplifiers at mm 
wavelengths have made short wavelength interferometer sensitivities 
competitive, CMB interferometers operating at 3 mm are being planned.

\noindent A consortium of major radio astronomy institutions across 
the world is
currently planning the world's next generation large radio telescope, the
Square Kilometre Array (SKA).  The telescope will have a total collecting
area of $10^{6}$~m$^{2}$ and will be capable of interferometric 
continuum imaging at metre and centimetre wavelengths down to at least 
3 cm and perhaps lower.  Although the SKA will operate at cm wavelengths,
where foreground discrete source confusion limits the achieveable
sensitivity for the imaging of the brightness temperature fluctuations in the 
background CMB sky, its extreme sensitivity due to its large collecting area
may make it possible to measure and subtract the point source contamination
at these relatively low operating frequencies, and thereby image the
low surface brightness CMB anisotropies on small angular scales.  We examine
the prospects for CMB observations with the SKA in the following sections.

\vfill\eject

\noindent {\bf DISCRETE SOURCE CONFUSION}

\noindent Discrete synchrotron sources in the sky are a foreground 
`contaminant' in images of CMB anisotropy.  They may be assumed to be 
Poisson random distributed on the sky with possibly some small degree 
of clustering. The clustering is indeed tiny: to our knowledge, the NVSS
was the first radio survey to detect 2-point angular correlation between 
radio sources on the sky [1].

\noindent The mean spectral index $\alpha$ ($S_{\nu} \sim \nu^{\alpha}$)
of extragalactic sources is about
$-0.7$ at metre wavelengths and flattens somewhat to about $-0.5$
at cm wavelengths: most discrete sources in the sky have lower flux
density at smaller wavelengths.  Therefore, in images of the sky 
made with a constant telescope beam, the flux density variations 
(in Jy beam$^{-1}$) owing to discrete extragalactic sources are expected 
to be smaller at shorter (mm) wavelengths as compared to longer (cm)
wavelengths.  In contrast, at wavelengths longward of about 1 mm, 
in the Rayleigh-Jeans part of the CMB spectrum, the expected variance 
in flux density owing to CMB temperature anisotropies 
in sky images made with a fixed telescope beam increases towards
shorter wavelengths as $\Delta S_{\nu} \sim \nu^{2}$.  
Searches for small angle anisotropy that have been done at cm
wavelengths, like the 3.4 cm ATCA search for arcmin scale CMB anisotropy [2],
were done in fields pre-selected to be relatively devoid of the stronger
point sources; however, the sensitivity was limited by confusion due
to weaker sources which could not be subtracted because of inadequate
sensitivity to these point sources.  For this reason, most planned
CMB telescopes are being designed for operation at mm wavelengths to
avoid the known extragalactic discrete source confusion that
dominates the CMB anisotropy at cm wavelengths.

\noindent All surveys of the sky for discrete sources, which have covered
significant parts of the celestial sphere, have been made at metre
and cm wavelengths.  Our catalogs of sources detected at mm wavelengths
have been made from observations of the mm flux densities of sources
detected in the long wavelength surveys.  Our knowledge of the source
counts in the mm sky is based on extrapolations of source populations
identified at cm wavelengths and may miss sources that are 
bright at mm wavelengths and undetected at cm wavelengths.  
Recent discoveries 
of high frequency peakers [3] and sources which have spectral index
$\alpha \sim 2$ at cm wavelengths [4] are indications that there may be
surprises for us when mm source counts and populations are directly
determined via mm surveys.  To sumarize, the movement of CMB anisotropy
search telescopes to short wavelengths certainly avoids the contamination
from most known source populations; however, the mm observations may now
have to confront new source populations and ill understood dust
contamination.  An alternate approach is to make small angle CMB surveys
at cm wavelengths and overcome confusion from the well studied cm source
populations via high sensitivity detections of these sources.

\noindent Relatively nearby sources 
have differential number counts $N(S) \sim S^{-2.5}$ as expected in a Euclidean
universe, weaker distant source counts deviate from this 
expectation and fall off 
as $N(S) \sim S^{-1.8}$ because of the space time structure of our expanding 
universe and source evolution. Normalized to the Euclidean expectation, an 
upturn is seen in the normalized differential source counts at $\mu$Jy 
flux densities because a new nearby population dominates the counts. The
differential source counts at $\mu$Jy flux density levels 
and at cm wavelengths are approximately \hfill\break
$$
N(S)~=~10^{2}~S_{\mu{\rm Jy}}^{-2.2}
~f_{GHz}^{-0.8}~{\rm arcmin}^{-2}~\mu{\rm Jy}^{-1}. \eqno(1)
$$ 
\noindent Any sky survey is made with a certain thermal noise sensitivity 
and a beam size.   There is a threshold flux density $S_{c}$ at which the 
integral source count $N(>S)$, at $S=S_{c}$ and within the beam area, is 
unity.  We expect one source on the average with flux density exceeding 
$S_{c}$ in any beam area.  
If a survey has a thermal noise that is less than or close 
to $S_{c}$, we would expect that sources in the sky which have a flux density 
well exceeding $S_{c}$ would stand out and could be reliably identified by 
the survey as `foreground sources'. These sources would occupy a 
small fraction 
of the image pixels (which I assume are roughly the size of the beam). 
If these identified sources are subtracted from the image, or if these 
few pixels that obviously contain sources are omitted from consideration, 
the weak sources with flux density less than $S_{c}$ will contribute an image 
rms of value approximately equal to the flux density threshold $S_{c}$. 

\noindent We would label any 
survey as being `sensitivity limited' if the thermal noise in that survey 
exceeds its $S_{c}$; a survey would be considered to be `confusion limited' 
if the thermal noise is less than $S_{c}$.
In images made with a beam of FWHM $\theta$ arcmin, the threshold
flux density is given by \hfil\break
$$
S_{c}~=~40~\theta^{1.7}~f_{GHz}^{-0.7}~\mu{\rm Jy}, \eqno(2)
$$
\noindent and the corresponding brightness sensitivity is \hfil\break
$$
\Delta T~=~14~\theta^{0.3}~f_{GHz}^{-2.7}~{\rm mK}. \eqno(3)
$$
\noindent $S_{c}$ and $\Delta T$ are, respectively, the confusion noise
limit and corresponding brightness sensitivity limit of the sky survey.

\noindent The relatively short baselines of the SKA would be useful for
imaging the CMB sky with high brightness sensitivity; we assume herein 
that CMB images are made with 1 arcmin resolution.  At 10 GHz, these images 
would have a confusion noise of 8~$\mu$Jy.  The corresponding brightness 
sensitivity limit is 30~$\mu$K.  

\noindent The CMB angular power spectrum is expected to have a band power
of $\sqrt{l(l+1)C_{l}/(2 \pi)} \sim 6~\mu$K on arcmin scales owing to the
SZE in cosmological clusters.  A SZE survey for distant clusters with the
SKA requires lowering the confusion noise via the subtraction of weak
sources in every beam area.

\noindent The proposed SKA is to have a sensitivity: 
$A_{eff}/T_{sys} = 2 \times 10^{4}$~m$^{2}$~K$^{-1}$; we assume a 10 percent
observing bandwidth.  The entire collecting
area (all baselines) would potentially be useful for detecting the discrete
source confusion in the fields.   We may, therefore, expect
the high resolution continuum images made using the full array 
to have a thermal noise of about 50~nJy with 1~hr integration time at 10 GHz.
Placing a 5-$\sigma$ threshold for the reliable detection of point soucres
in the field, we may expect sources with flux density exceeding about 
250~nJy to be subtracted.  There would be, on the average, about $10^{2}$
such sources in every arcmin area at 10~GHz.

\noindent If we assume that all discrete sources above a lower flux density
limit of $S_m~\mu$Jy are subtracted from the sky images, the residual
confusion rms is \hfill\break
$$
\Delta S~=~8~\theta~f_{GHz}^{-0.4}~S_m^{0.4}~\mu{\rm Jy}, \eqno(4)
$$
\noindent and the corresponding brightness sensitivity is
$$
\Delta T~=~3~\theta^{-1}~f_{GHz}^{-2.4}~S_m^{0.4}~{\rm mK}. \eqno(5)
$$
\noindent Assuming that foreground sources above $S_m \sim 250$~nJy
are successfully subtracted, the residual confusion in arcmin resolution
images may be expected to be as low as 6~$\mu$K
at 10~GHz.

\noindent It appears that the SKA could usefully image the background CMB at
a frequency $\gsim 10$ GHz, with arcmin resolution, and with a confusion
noise limit less than 6 $\mu$K.  If the thermal noise is to be less than 
the residual confusion noise, say a tenth of the confusion noise, 
a quarter of the SKA baselines are to be useful in making the arcmin scale
CMB survey images.

\noindent {\bf INTERFEROMETRIC CMB IMAGING WITH THE SKA}

\noindent Interferometers are the instruments of choice for precision
measurements of extremely weak sky signals like the CMB anisotropies 
because they have much lower systematic errors.  The SKA, being an
interferometric imaging telescope, would have substantial advantages
of lower systematics.  The lower element sensitivity, owing to the
smaller bandwidths at the longer operating wavelengths, would be
offset by the sensitivity gain obtained by the extremely large number
of antenna and receiver elements.  Additionally, at the longer cm
wavelengths, system temperatures are lower owing to lower receiver
noise temperatures, lower losses in signal transmission, lower
atmospheric emission temperature.  

\noindent The inherent stability of
interferometers  implies that highly precise calibration, including
polarization calibration, is possible.  In some of the SKA configuration 
solutions, the stations have a large number ($n$) of elements in
compact arrays.  These arrays have of order $n^{2}$ longer baselines 
that will resolve the CMB but will measure the discrete sources with 
high accuracy.  Since the same elements measure the long and short 
baselines simultaneously, the subtraction of self calibrated intensity 
and polarization should be accurate to better than 1 part in 1,000.  

\noindent Finally each SKA patch can make an independent observation, 
and these could either be used to decrease noise in one patch of sky, 
or to observe patches over a larger area to increase the survey area, 
in searches for cosmological clusters, or to reduce cosmic variance, 
in statistical measurements of CMB anisotropy.

\noindent {\bf REFERENCES}

\noindent [1]   C. Blake, and J. Wall, ``Measurement of the angular 
correlation function of radio galaxies from the NRAO VLA Sky Survey,''
{\it Mon. Not. R. ast. Soc.}, vol. 329, pp. L37-L41, 2002. \hfill\break
\noindent [2]   R. Subrahmanyan, M.J. Kesteven, R.D. Ekers, M. Sinclair,
J. Silk, ``An Australia Telescope survey for CMB anisotropies,
'' {\it Mon. Not. R. ast. Soc.}, vol. 315, pp. 808-822, 2000. \hfill\break
\noindent [3]   D. Dallacasa, C. Stanghellini, M. Centonza, R. Fanti,
``High frequency peakers, I. the bright sample'',
{\it Astr. Astrop.}, vol. 363, p. 887-900, 2000. \hfill\break
\noindent [4]   A.C. Edge, G. Pooley, M. Jones, K. Grainge, R. Saunders,
``GPS sources with High Peak Frequencies'', in {\it Radio Emission from
Galactic and Extragalactic Compact Sources}, ASP Conf. Ser., Vol. 144,
IAU Colloquium 164, eds. J.A. Zensus, G.B. Taylor, and J.M. Wrobel, p. 187,
1998. \hfill\break

\vfill\eject
\end